\documentclass[aps,prl,twocolumn,showpacs,preprintnumbers,amsmath,amssymb]{revtex4-1}

\usepackage{graphicx}
\usepackage{dcolumn}
\usepackage{bm}

\begin{document}

\title{Optimal Control of \textit{Lantana camara}: An Entropy-Based Sustainable Strategy}
\author{Shyam Kumar}
\author{Preet Mishra}
\author{R K Brojen Singh}\email{brojen@jnu.ac.in}
\affiliation{School of Computational and Integrative Sciences, Jawaharlal Nehru University, New Delhi, India}

\begin{abstract}
{\noindent}Framing control policies to mitigate the impact of invasive plants on indigenous biodiversity within the Sustainable Development Goals (SDG) framework is the primary objective of this work. Using reported ecological dynamics of the invasive species \textit{Lantana camara}, we develop a minimal three-species network model, where each node follows generalized Lotka-Volterra (GLV) dynamical equations. Employing Lie algebra and network control theory, we establish the model's controllability and accessibility criteria. Through nonlinear optimization programming, we derive sustainable policies for controlling abundances of \textit{Lantana camara}. We also have used Shannon entropy as an indicator to assess the sustainability of these optimal policies. The analysis of the sensitivity measured using this technique reveals that the control strategy is critically dependent on the ratio of the intrinsic growth rates of the \textit{Lantana camara} and the control plant. Thus, we get a modular algorithmic decision support mechanism for designing control policies to manage \textit{Lantana camara} abundances. \\


\end{abstract}

\maketitle

\section{Introduction}
{\noindent}Strategic control of invasive plants in an ecosystem is a global issue as they are one of the major causes of biodiversity loss, environmental damage and threat to endangered native species. In this work We deal with controlling the invasive species \textit{Lantana camara} which is one of the main drivers of imbalance between local-global plant diversity\cite{Sivakumar2018,Ramesh,Garkoti2021,Kohli2006,Hiremath2018}. This complex process of invasion leads to a "state of dominance by the invasive species" in the ecosystem which we have termed as \textit{diseased state}. We also know \textit{Lantana camara} promotes the growth of deleterious microbes that decrease the population of beneficial microbes in the soil \cite{Garkoti2021, Fan2010} and implicitly hinder the growth of other plants. The main goal of this work is to propose a model and analyse it to guide the diseased ecosystem to a possible desired healthy state \cite{Montoya,Jones,Singh,Weidlich} in a sustainable manner as suggested in the United Nations Sustainable Development Goals (UN-SDG)\cite{un} using suitable external inputs. We have thus defined two-fold objectives, first we want to manage the abundance of \textit{Lantana camara} population and second, to give a possible mathematical framework to address important issues of sustainability to control the invasive plants in general. We derive our theoretical model systematically by following some of the basic ecosystem management rules.\\

{\noindent}The basic ecological observations which are incorporated in our model framework are the following. For our control objectives we introduce a Control-plant(C-plant) into the ecosystem which helps us control the abundance of \textit{Lantana camara} and also increases the soil microbe abundances.  There are studies showing ways to use the plant-plant interaction to control the abundances of plants that live within a spatial distance threshold from each other in the ecosystem \cite{Joshi1991,Sakai1954}. This interaction also has its own benefits as certain plants are known to improve soil quality by increasing and promoting the growth of good microbes in the soil\cite{SinghHP,Hamilton,Wardle}. It has been found in certain field studies\cite{Joshi1991} that some non-invasive plants like \textit{Ipomoea staphylina}, \textit{Mimosa pudica} (Mimosa has also added benefits of being a leguminous plant) can act as being control plants, which further provides a practical basis to our theoretical modelling and analysis. From these basic important observations, we propose a minimal model, which consists of these two fundamental interactions i.e. plant-plant interactions and the plant-microbe interactions, and analyse this model for possible sustainable control strategies of invasive species \textit{Lantana camara}. The dynamics of each nodal variable is defined by generalized Lotka-Volterra (GLV) model. The control strategies in the model are incorporated using nonlinear control theory \cite{Angulo2019,Liu2016}. This is a basic minimal model specially designed to incorporate control theoretical technique and can be extended to more general interactions.\\

{\noindent}The behavior of such ecological systems can be well studied within the framework of population dynamics, which studies how the population states of species of the system change over time due to interactions within individuals of the same species and with individuals of other species present in the ecosystem. If we know the dynamical behavior of the system, a desired state can be reached from a specified state through external actions i.e. by application of a suitable control strategy \cite{Angulo2019,Liu2016,Liu2011,Conte1999,Brockett2015,Brockett1973,Nijmeijer1990}.\\

\begin{figure*}[htbp]
    \centering
    \includegraphics[scale=0.2]{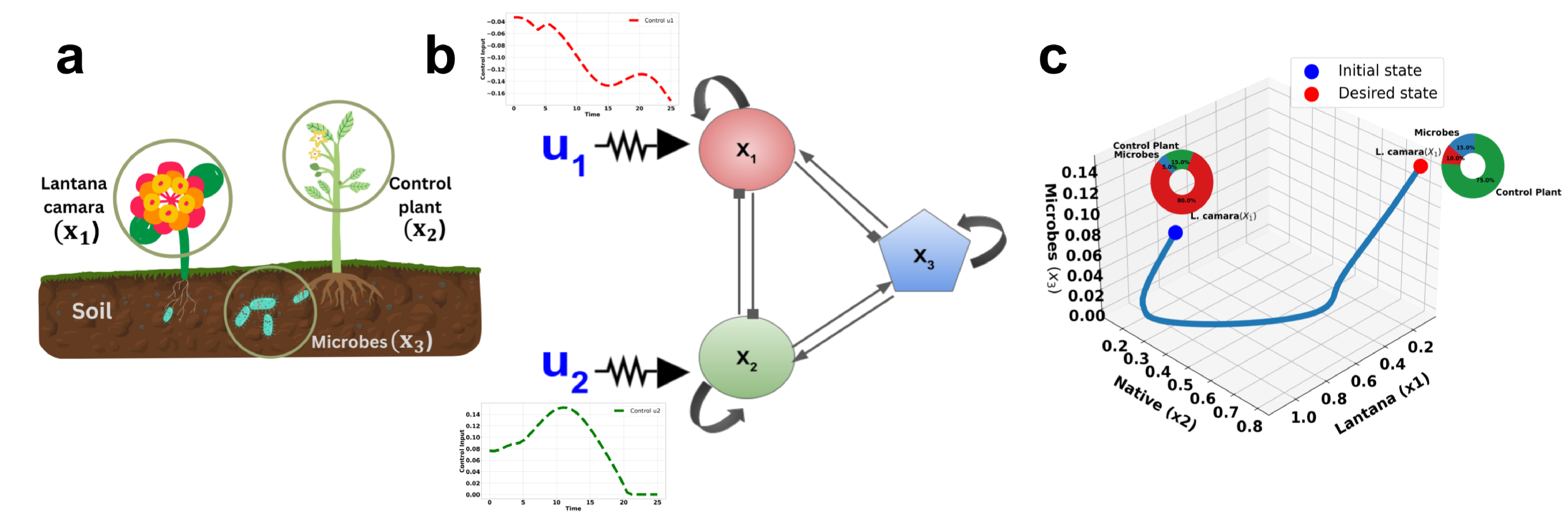}
    \caption{ Schematic of the ecological system, its modeling, and controlled state evolution.\textbf{(a)} Shows the eco-system of the three interacting species :\textit{Lantana camara}, control plants, and microbes.\textbf{(b)} A network representation of GLV type interactions among ecological agents. The abundances of \textit{Lantana camara} \(x_1(t)\), control plants \(x_2(t)\), and microbes \(x_3(t)\)  as nodes in the network. Control inputs \(u_1(t)\) (reducing \textit{Lantana camara}) and \(u_2(t)\) (enhancing control plants) act on \(x_1\) and \(x_2\) respectively, with inset plots indicating \(u_j(t)\) .\textbf{(c)} A state-space plot displays the system’s trajectory from the initial (blue) to desired (red) state, alongside pie charts showing species proportions aimed at reducing \textit{Lantana camara} while boosting control plants and microbes.}  
    \label{init9}
\end{figure*}

\section{Model Description}
{\noindent}Components of an ecosystem such as plants, animals, soil microbes, etc. interact in a complex manner as shown in Figure \ref{init9}(a) and can be represented by a network as shown in Figure \ref{init9}(b). In such networks defined by a graph $G(V,E)$, ecological agents are represented by the nodes (\textbf{V}) and interactions by edges (\textbf{E}) among them \cite{Angulo2019,Cohen1968,Roberts1978}. In our study, we represent the dynamics of the individual nodes by generalized Lotka-Volterra model(GLV) \cite{May1975}. Once we construct the network dynamical system, we then ask how we can steer the system in a finite amount of time to achieve a certain desired state by suitable input signals to the chosen nodes (ecological agents) of the network as shown schematically in Figure \ref{init9}(c). For the analysis and solution to this situation, network theory-based controllability \cite{Angulo2019,Liu2011,Liu2016} predictions have been systematically implemented in this work.\\

{\noindent}The minimal ecological model we propose in this work is three nodes network, whose individual dynamics is governed by nonlinear GLV \cite{Goel,May1975,Mambuca}, with the interacting edges built from the experimental observations\cite{Sivakumar2018,Ramesh,Garkoti2021}. We define the state vector $\textbf{X}=[x_1,x_2,x_3]^T$ where we represent the abundances of \emph{Lantana camara} ($x_1$), control-plants ($x_2$) and soil microbes ($x_3$).We have considered the interaction strength matrix (the adjacency matrix) A $[A]_{3\times 3}$ with $0<|A_{ij}|<1$ giving the relative strength of interactions between the nodes(the participating ecological agents ) of the network in the Generalized Lotka Volterra (GLV) model. Thus for this model, the uncontrolled dynamical equations are given by:
\begin{widetext}
\begin{eqnarray}
\label{glv}
\frac{d\bf{X}}{dt}=F(\bf{X})=D[\bf{X}]\left[A\bf{X} + r\right],~~A=\left[\begin{matrix}-A_{11}&-A_{12}&A_{13}\\
-A_{21}&-A_{22}&A_{23}\\
-A_{31}&A_{32}&-A_{33}
\end{matrix}\right],~~
D[X]=\left[\begin{matrix}x_{1}&0&0\\
0&x_{2}&0\\
0&0&x_{3}
\end{matrix}\right],~~
r=\left[\begin{matrix}r_1\\r_2\\r_3\end{matrix}\right]
\end{eqnarray}
\end{widetext}
where, $r_i;i=1,2,3$ represent the intrinsic growth rates adjusted for carrying capacities of the ecological agents. These equations represent the system without the control signals. The biomass proportions are normalized such that the total biomass sums to 1, with plants contributing 85\% and microbes 15\% to the total. This normalization ensures that the model accounts for the relative biomass contributions while maintaining ecological relevance.\\

{\noindent}In our model, we imposed two control signals $u_1(t)$ and $u_2(t)$ on the abundance variables of \textit{Lantana camara}($x_1$)  and C-plant ($x_2$) respectively. The choice is based on the fact that $x_1$ and $x_2$ are the driver nodes which are identified by using the maximum matching algorithm to our system \cite{Angulo2019}. This algorithm can be applied to identify suitable driver nodes at which control inputs can be applied \cite{Angulo2019,Brockett1973}. Now, after control input signals are incorporated in Eq.\eqref{glv}, we have the following control equation:
\begin{equation}
\frac{d\textbf{X}}{dt}=F(\textbf{X})+\sum_{j=1}^{2}f_j u_j(t); \quad  f_1=\left[\begin{matrix}1\\0\\0\end{matrix}\right],~~
f_2=\left[\begin{matrix}0\\1\\0\end{matrix}\right]
\label{auto}
\end{equation}
 where, it was assumed that the susceptibilities ($f_{j}$) of the species to input are constant and equal to one. \\

\noindent \textbf{Controllability and accessibility:}
The general controllability problem can be defined as steering the systems to reach the desired \textit{healthy state} $X_D$ from a given initial \textit{disease state} $X_{0}$ in a finite time with appropriate input signals $u_{j}(t)$. To steer the system to the desired state, we need to check the desired state's accessibility and hence prove its controllability \cite{ Brockett1973,Conte1999, Nijmeijer1990}. For checking the accessibility, we have to calculate the matrix $\mathcal{C}$ given by,
\begin{eqnarray}
\label{matrix}
\mathcal{C}=[f_1, f_2, [f_1,F],[f_2,F],[f_1,[f_1,F]], [f_2,[f_2,F]]... ]
\end{eqnarray}
where, $[...]$ is the commutator, such that,
\begin{eqnarray}
[f, F](x)=\frac{\partial F}{\partial x} f(x)-\frac{\partial f}{\partial x} F(x)
\end{eqnarray}
are the Lie brackets of $F$ and $f$ vector fields. The ecological network dynamics\cite{Angulo2019,Conte1999} with control input signals $u_j$ is said to be controllable to reach from $X_0$ to $X_D$ if and only if the controllable matrix given by \eqref{matrix} has full rank, which is called Kalman's controllability rank condition \cite{KalmanR}, given by,
\begin{eqnarray}
\label{Kalman}
rank[\mathcal{C}]=n
\end{eqnarray}
For equation \ref{auto} the system is said to be accessible if it does not have any autonomous elements. Autonomous element are internal variables of the system and are not affected by the system control (details of proof can be seen in \cite{Angulo2019,Conte1999}).\\

{\noindent}Further we apply Shannon entropy to evaluate the degree of the sustainability of the control policies of the model. Shannon entropy is a fundamental information measure used to quantify the diversity in a distribution \cite{Jianhua}. It is mathematically defined as: $H = -\sum_{i=1}^{n} p_i \log(p_i),$ where \( p_i \) represents the proportion of data in the \( i \)-th bin, and \( n \) is the total number of bins \cite{Shannon}. The entropy reaches its minimum value, \( H = 0 \), when all the data is concentrated in a single bin (\( p_i = 1 \) for one bin and \( p_j = 0 \) for all \( j \neq i \)), indicating no diversity or uncertainty. Conversely, the entropy is maximized when the data is evenly distributed across all bins (\( p_i = \frac{1}{n} \) for all \( i \)), in which case \( H = \log(n) \), represents maximum diversity. For example, with six bins, the maximum entropy is \( H = \log(6) \approx 1.7918 \).\\

{\noindent}We identify sustainability with the idea that any action we take must not cause further depletion of resources. For example, if an action plan is designed such that one type of action is taken over a long period, it results in a cost-effective but not sustainable policy. Thus, we used Shannon entropy as an indicator of the degree of uniformity in the actions taken, i.e., we need to classify actions that use the maximum possible combinations in the control input space. Policies with higher entropy indicate a more robust application of interventions, such as non-monotone cutting of Lantana (\( u_1 \)) and planting of control plant species (\( u_2 \)), which ensures ecological stability and sustainability.\\

{\noindent}We design our model predictive control (MPC) for our problem of controlling the population of the Lantana as defined by the dynamical system. This algorithm aims to solve the impulsive control problem. The problem of finding a control input sequence that is time-optimal is a variational problem requiring an optimization of a cost function under the constraints of given control inputs. We developed a cost function given as:\begin{widetext}
    \begin{equation}
        J= C_{\text{control}} \cdot \left(u_1^2 + u_2^2\right) + C_{\text{biodiversity}} \cdot \left((x_1 - x_{1d})^2 + (x_2 - x_{2d})^2 + (x_3 - x_{3d})^2\right) 
        \label{cfunc}
    \end{equation}
\end{widetext}  that balances two critical objectives: $C_{\text{control}}$, which penalizes aggressive control efforts ($u_1, u_2$), and $C_{\text{biodiversity}}$, which penalizes deviations from the desired ecological state ($x_1, x_2, x_3$). The optimization has been done by using a two parameters cost function. We have used a standard Non-Linear Programming Solver \texttt{CASADI} available in \texttt{Python}.

\section{Results}

\begin{figure}[htbp]
    \centering
    \includegraphics[scale=0.295]{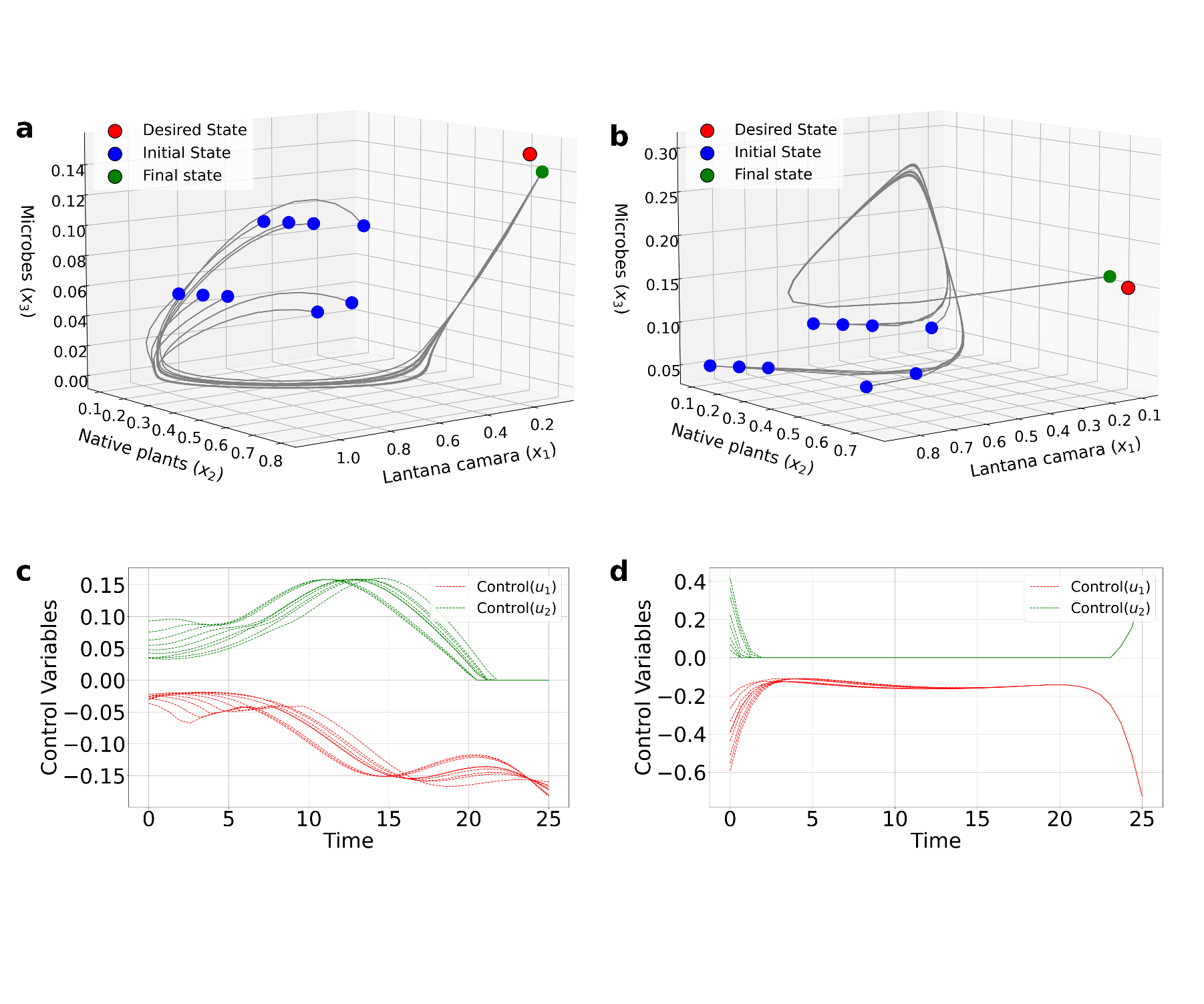}
    \caption{Control dynamics of the system across various initial species abundances. 
Growth rates: $r_{1} = 0.85$, $r_{2} = 0.4$, $r_{3} = 0.3$. 
Desired state: $(x_1^d, x_2^d, x_3^d) = (0.1, 0.75, 0.15)$. 
Interaction matrix: 
$a_{11} = 0.6$, $a_{12} = 0.6$, $a_{13} = 0.5$, 
$a_{21} = 0.6$, $a_{22} = 0.3$, $a_{23} = 0.5$, 
$a_{31} = 0.7$, $a_{32} = 0.6$, $a_{33} = 0.7$. 
Cost parameters: 
(a) and (c): Control cost $C_{\text{control}} = 1000.0$, 
biodiversity penalty $C_{\text{biodiversity}} = 1.0$. 
(b) and (d): Control cost $C_{\text{control}} = 1000.0$, 
biodiversity penalty $C_{\text{biodiversity}} = 1000.0$.}
 
    \label{fig:control2}
\end{figure}

{\noindent}We provide solutions to the optimal control problem of the management of abundances of the invasive plant \textit{Lantana camara}. We also analyzed the qualitative and quantitative behavior of these calculated optimal control policies which can drive the system from any initial condition to a given desired state. The main importance of the work is, first we designed a GLV network-based dynamical control systems model using Lie algebraic approaches. Secondly, by using the ideas of information theory to the model, where, we used Shannon entropy as a measure of the degree of sustainability of the control policies adhering to the paradigms of Sustainable Development Goals (SDG). \\
\begin{figure*}[htbp]
    \centering
    \includegraphics[scale=0.47]{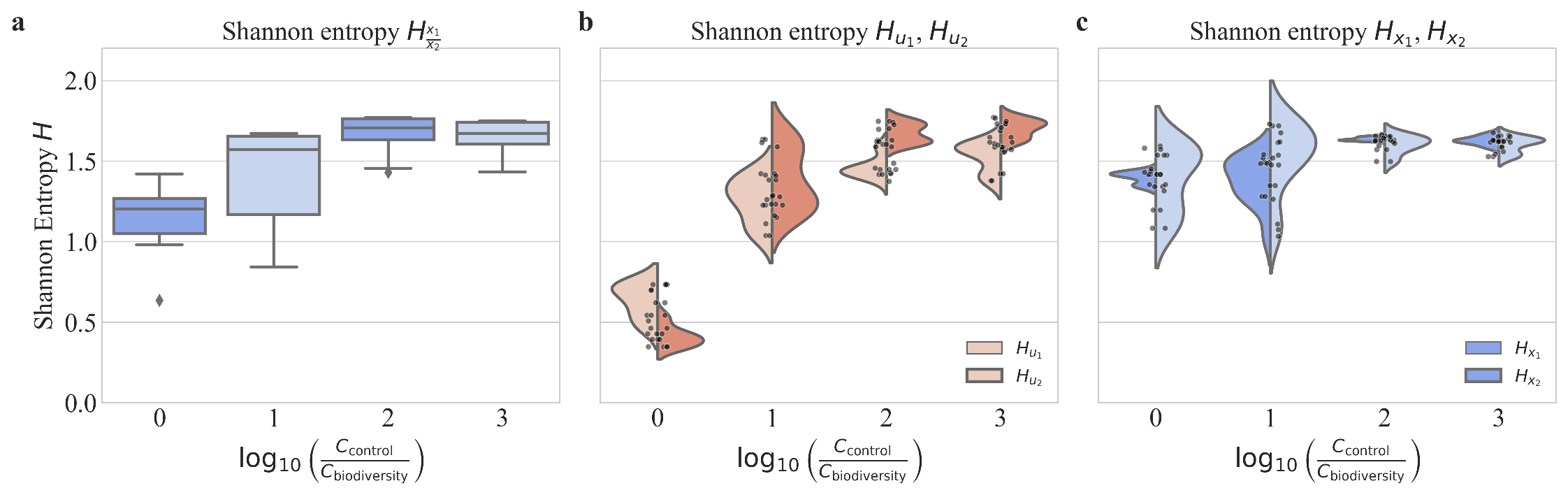}
    \caption{Shannon entropy \( H \) for evaluating the sustainability of control policies. (a) Box plot of \( H \) against \( \log(c_{\text{control}}/c_{\text{biodiversity}}) \) for the ratio of species abundances $\frac{x_1}{x_2}$,this gives the information about the distribution, in time, of the values of $\frac{x_1}{x_2}$ This indicates how this ratio behaves in time when initially it is $\frac{x_1}{x_2}>1$ and crosses over to $\frac{x_1}{x_2}<1$, across a set of varied initial conditions. 
(b)  \( H_{u_1} \) and \( H_{u_2} \) from the distribution of the control signals in time for the same set of initial conditions as in (a). 
(c) \( H_{x_1} \) and \( H_{x_2} \) for species abundance distribution in time for the same set of initial conditions as in (a), 
reflecting control effectiveness and biodiversity outcomes.}  
    \label{fig:shannon}
\end{figure*} 

{\noindent}Further, we have also verified the accessibility and controllability of the system's dynamics. For checking accessibility, the Lie Algebra Rank Condition (LARC) was applied by constructing the accessibility matrix $\mathcal{A}(x)$, which includes higher-order Lie brackets of the vector fields. The rank of $\mathcal{A}(x)$ was determined to be 3, confirming that the system satisfies the LARC. This result indicates that the system dynamics allow sufficient freedom to explore the state space under the given controls. Based on this criteria, the controllability matrix $\mathcal{C}(x)$ was constructed by incorporating the control vector fields and their Lie brackets with the drift term. The rank of $\mathcal{C}(x)$ was also found to be 3, matching the dimension of the state space. This demonstrates that the system is fully controllable, ensuring that the control inputs can steer the system from any initial state to any desired state within a finite time interval. Combining together, these results validate that the designed control policies are both theoretically sound and practically feasible for managing invasive species while ensuring ecological sustainability. Detailed calculations for accessibility and controllability conditions are provided in the Appendix.\\

{\noindent}We show in Fig.\ref{fig:control2} how our predictive control model works and  how to achieve optimal control policies that can drive the system from any initial condition to a given desired state within a finite time interval. In Figure \ref{fig:control2}(a,b) we show the robustness of the control program by applying control signals to the dynamics initiated from various initial conditions. We also show that with suitable choice of the system's parameter values, we can drive the system to the desired state of abundances of Lantana camara within a finite time span. On changing the two parameters of the cost function, i.e., the control cost penalty parameter ($C_{Control}$) and the Bio-Diversity Penalty parameter ($C_{biodiversity}$) used in the optimization technique, we observed the different trajectories available Figure \ref{fig:control2} c,d. Selection of the correct policy requires another sieve, which we now elucidate. When designing control strategies for managing the invasive species \textit{Lantana camara}, optimizing policies solely based on a fixed cost function is insufficient to ensure ecological sustainability. To address this limitation, we introduced Shannon entropy as an additional metric to evaluate and select control policies. Shannon entropy quantifies the diversity and balance of control efforts, providing insights into how evenly the management inputs are distributed over time. Figure \ref{fig:control2} c,d illustrates various optimal control policies, each with distinct ecological implications. For example, the policy in Figure \ref{fig:control2}c employs a non-monotone application of control inputs—cutting \textit{Lantana camara} ($u_1$) and planting control plant species ($u_2$)—throughout the management period. This approach promotes a gradual replacement of \textit{Lantana camara} with control plants, fostering ecological balance while minimizing disruptions to soil health, microbial communities, and the broader environment. In contrast, the policy in Figure \ref{fig:control2}d adopts a more aggressive strategy, concentrating intensive cutting and planting efforts at the beginning and end of the management period, with prolonged intervals of inaction in between. Such abrupt removal of \textit{Lantana} without consistent planting risks leaving the land vulnerable to degradation, leading to soil erosion, nutrient depletion, microbial imbalance, and other cascading ecological issues. While this strategy may seem cost-effective, it compromises long-term stability and sustainability.\\

{\noindent}We report the impact of changes in the parameters of the cost function on characterization of the feasible optimal control policies. Using the defined cost function $J$ in \eqref{cfunc}, and keeping the interaction matrix fixed, we computed the optimal control trajectories for various parameter pairs: $(C_{\text{control}}, C_{\text{biodiversity}}) = (1000, 1000), (1000, 100), (1000, 10), (1000, 1)$, across multiple initial conditions. For each trajectory, we have used six bins for discretization (We also found that on increasing the number of bins the distribution did not change in shape). The Shannon entropy ($H$) was calculated for the time series of $x_1$, $x_2$, $u_1$, $u_2$, and the ratio of the Lantana and control plant abundances $\frac{x_1}{x_2}$. As can be seen in the distribution of the control inputs shown in figure \ref{fig:shannon}b,  in the case of the example trajectory in \ref{fig:control2}(b,d) corresponds to the ratio of $log_{10}( \frac{C_{control}}{C_{biodiversity}}) = 0$ having a lower value of Shannon entropy as compared to the trajectory in \ref{fig:control2}(a,c) corresponding to the ratio of $log_{10}( \frac{C_{control}}{C_{biodiversity}}) = 3$ which has a higher entropy value. From these results observed and analysis, we report that there is a meaningful way of designing sustainable policy for a desirable ecological state by the way of maximizing the Shannon entropy as the basis of selection of the action plan to be taken to control the abundances. This trend encourages a more uniform distribution of control efforts over time, as evident with higher entropy values. Policies with higher entropy condition align with ecological sustainability principles by ensuring continuous and evenly distributed management efforts, ultimately promote a balanced and desirable resilient ecosystem.\\
\begin{figure}[htbp]
    \centering
    \includegraphics[scale=0.4]{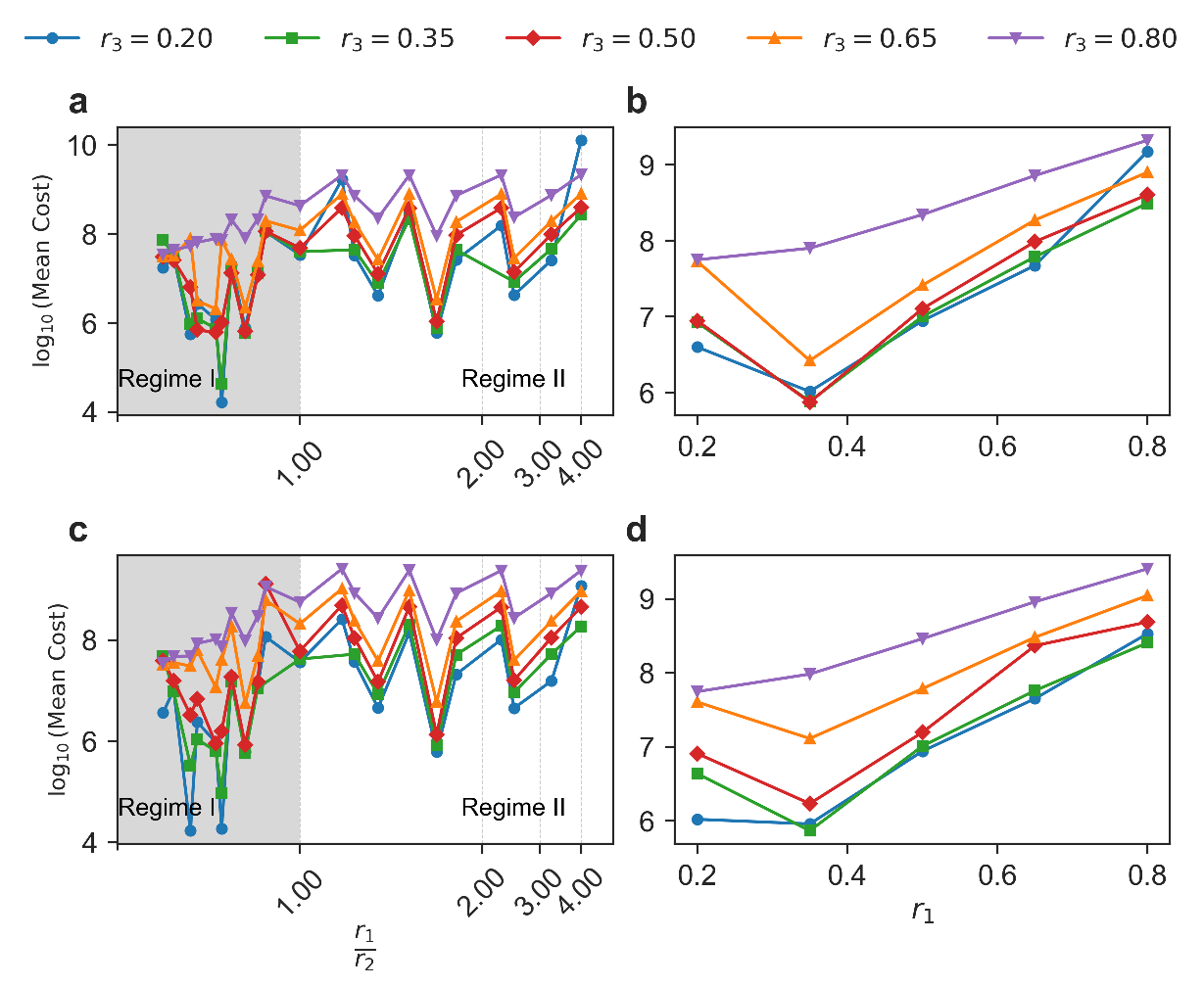}
    \caption{Sensitivity analysis: Mean control costs show an increasing trend with an increase in the growth rate of \textit{Lantana}. (a) and (b) correspond to the initial condition of high abundance of Lantana with the representative figures as (0.85, 0.1, 0.05). Figures (c) and (d) show results for a moderate abundance of Lantana with the  initial condition (0.6, 0.3, 0.1). The interaction matrix is fixed as:
    $a_{11} = 0.3  ,a_{12} = 0.1 , a_{13} = 0.4 , a_{21} = 0.8  ,a_{22} = 0.2  ,a_{23} = 0.5,   a_{31} = 0.6 , a_{32} = 0.7 , a_{33} = 0.7$. The mean is taken over for different values of parameter $r_2$.} 
    \label{fig:Sensitivity}
\end{figure}

{\noindent}We study sequentially how sensitive are the proposed control policies to the changes in model parameter values. There are two types of parameters regulating our proposed model, firstly, we have the dynamical systems parameters, and, secondly, we have the parameters in the optimization of control cost function. We check and systematically analyzed the sensitivity of the control cost to both types of parameters. Since the precise values of model parameters are unknown, we analyzed the trends of control costs by uniformly varying intrinsic growth rates (\(r_1, r_2, r_3\)) across the range \([0.2, 0.8]\), generating 125 tuples while keeping the values of the elements of the interaction matrix fixed. Simulations were conducted for two initial conditions: \textit{Case I} as shown in Figure~\ref{fig:Sensitivity}(a-b), where, the abundance of the Lantana camara is high over the landscape (\(x_1, x_2, x_3 = 0.85, 0.1, 0.05\)), and \textit{Case II}, where its abundance is moderate (\(x_1, x_2, x_3 = 0.6, 0.3, 0.1\)) as shown in the Figure~\ref{fig:Sensitivity}(c-d). The analysis revealed two distinct regimes based on the ratio of Lantana camara’s growth rate (\(r_1\)) to that of the control plants (\(r_2\)) which matches with the intuitive idea that we can use only certain plants to outgrow Lantana camara. In \textit{Regime I}($\frac{r_1}{r_2}<1$) , where, the growth rate of Lantana camara is lower than that of control plants, control costs are relatively low. In contrast, \textit{Regime II} (($\frac{r_1}{r_2}>1$)), where, Lantana camara outcompetes control plants with higher growth rates, incurs significantly higher management costs, as evident in Figures~\ref{fig:Sensitivity}(a) and \ref{fig:Sensitivity}(c). These trends underscore the critical role of competition dynamics, with higher costs arising when Lantana camara dominates resource consumption and growth. Figures~\ref{fig:Sensitivity}(b) and \ref{fig:Sensitivity}(d) further illustrate that the behaviour of the mean (averaged over various values of $r_2$) control costs for scenarios of high and moderate Lantana camara abundances. The results clearly show that as \(r_1\) increases, the management costs consistently rise. While the system shows limited sensitivity to the small variations within each regime, the absolute cost strongly depends on the relative growth rates and the initial abundance of the Lantana camara. These findings significantly highlight that the effective management of invasive species like Lantana camara requires addressing growth dynamics and competition, as higher growth rates and competitive dominance significantly escalate management efforts to restore ecological balance.\\

\noindent \textit{In conclusion}, our study on optimal control strategies for managing invasive plant, \textit{Lantana camara}, highlights the crucial role of \textit{Lantana camara}'s growth rate and competition with control plants in determining control costs, with two regimes identified as, low costs when Lantana camara's growth rate is lower than control plants, and, high costs when Lantana outcompetes them. By integrating Shannon entropy as a metric for sustainability, we ensure that control policies are evenly distributed over time, promoting long-term ecological stability and minimizing risks like soil erosion and microbial imbalance. Sensitivity analysis shows that control costs are influenced by both intrinsic growth rates and initial conditions, emphasizing the need for adaptive, balanced management strategies. Ultimately, our approach ensures that control policies, not only minimize the costs, but also foster ecological sustainability and resilience in the restoration of ecosystems affected by invasive species.\\

\noindent\textbf{Author Contributions:}
SK, PM, and RKBS conceptualized the model. SK and PM performed the mathematical analysis and simulations. RKBS supervised the work. All of the authors contributed to writing the manuscript.\\

\noindent\textbf{Acknowledgements:}
The authors would like to thank A. Langlen Chanu, Jyoti Bhadana, S.C.C. Vyom and Rubi Jain for valuable discussions. PM and SK would like to acknowledge JNU and UGC for financial support during the work. RKBS acknowledges the financial support from the DBT-BIC.\\

\noindent\textbf{Conflict of Interest:}
The authors declare that there is no conflict of interest.\\

\noindent\textbf{Data and Code Availability:}
The data generated in this study can be reproduced using the outlined methodology. The numerical codes used to produce the results are available upon reasonable request.\\

\end{document}